\def\sqig{$\sim\,$} \def\etal{et\,al.} 
\def\up#1{$^{\mbox{{\scriptsize #1}}}$} 
\def\pten#1{$\times10^{#1}$}
\def\minone{$^{-1}$} \def\mintwo{$^{-2}$}\def\kev{ke\kern-.1em V}
\def\appro{$\approx\,$}\def\rwd{$R_{\rm wd}$}
\def\xte{{\sl RXTE\/}}
\documentstyle{mn}
\title[GK~Per in outburst]
{On the magnetic accretor GK~Persei in outburst}
\author[C. Hellier, S. Harmer \&\ A.\,P.\,Beardmore]
{Coel Hellier, Sean Harmer and A.\,P.\,Beardmore\\
School of Chemistry and Physics, Keele University, Staffordshire, ST5 5BG}
\begin{document}
\maketitle
\begin{abstract}
\xte\ made 5 X-ray observations of the magnetic accretor GK~Per
during its 1996 outburst, recording a count rate of ten times the
quiescent level. The 351-s spin pulse shows a deep, nearly sinusoidal
modulation, in contrast to the weaker, double-humped profile of 
quiescence. The spectrum shows absorption 
increased by two orders of magnitude over quiescence.  We explain
these differences in terms of the changing accretion geometry as
the outbursting disc forces the magnetosphere inwards, and discuss
the 5000-s X-ray QPOs seen during GK~Per's outbursts.
\end{abstract}
 
\section{Introduction}
The dwarf-nova outbursts of cataclysmic variables are thought to be
caused by accretion-disc instabilities (e.g.\ Osaki 1996; Lasota 2001).
In principle, such outbursts can still occur when the inner disc is
truncated by the magnetic field of the white dwarf, as in the
intermediate polar subclass (e.g.\ Angelini \&\ Verbunt 1989).  GK~Per
is an exemplar of this: its long, 2-d orbit and short, 351-s,
white-dwarf spin period lead to a large disc surrounding a relatively
small magnetosphere. Its outbursts, which last for \appro 50 d
and recur every \appro 3 y (Simon 2002), can thus be modelled using a
disc-instability code with the inner disc missing (Kim, Wheeler \&\
Mineshige 1992; Yi \etal\ 1992).

Outbursts in some other intermediate polars may also be the result of disc
instabilities, for example in XY~Ari (Hellier, Mukai \&\ Beardmore
1997), YY~Dra (Szkody \etal\ 2002), HT~Cam (Ishioka \etal\ 2002), and
possibly EX~Hya (Hellier \etal\ 2000). It is likely, though, that
short-lived, low-amplitude `flares' seen in TV~Col and V1223~Sgr are
caused by something else, such as mass-transfer events (Hellier \&\
Buckley 1993).

Of the above systems, X-ray observations in outburst have been
obtained for GK~Per, XY~Ari, YY~Dra and EX~Hya.  Outburst observations of
GK~Per include {\sl EXOSAT\/} coverage of its 1983 outburst (Watson,
King \&\ Osborne 1985), {\sl Ginga\/} coverage of its 1989 outburst
(Ishida \etal\ 1992), \xte\ coverage of its 1996 outburst, and,
most recently, {\sl Chandra,} {\sl XMM-Newton\/} and \xte\ 
coverage of the 2002 outburst (Mauche 2003). 

We report here on the \xte\ observations
of the 1996 outburst.  In particular we address the issue of why the
351-s pulsation is strong and single-peaked in outburst (Watson \etal\
1985) but much weaker and double-peaked in quiescence
(Norton, Watson \&\ King 1988; Ishida \etal\ 1992).

\section{Observations and results}
\xte\ made five observations of the 1996 outburst of GK~Per, lasting
$\approx$\,9 h each, spaced over 41 d (see Fig.~1). We make use of
the PCA data extracted from the top xenon layer of PCUs 0, 1 \& 2 in
the energy range 2--15\,\kev, and with the background estimated using
{\sc pcabackest} v2.1e.

\begin{figure}\vspace*{83mm}     % Fig 1 
\caption{The visual lightcurve of the 1996 outburst (circles), compiled by
the AAVSO (Mattei 2003). Also shown are the times and count rates of
the 5 \xte\ observations (squares); the line marks the expected \xte\ 
count rate in quiescence, predicted from the quiescent {\sl Ginga\/}
count rate.}
\includegraphics{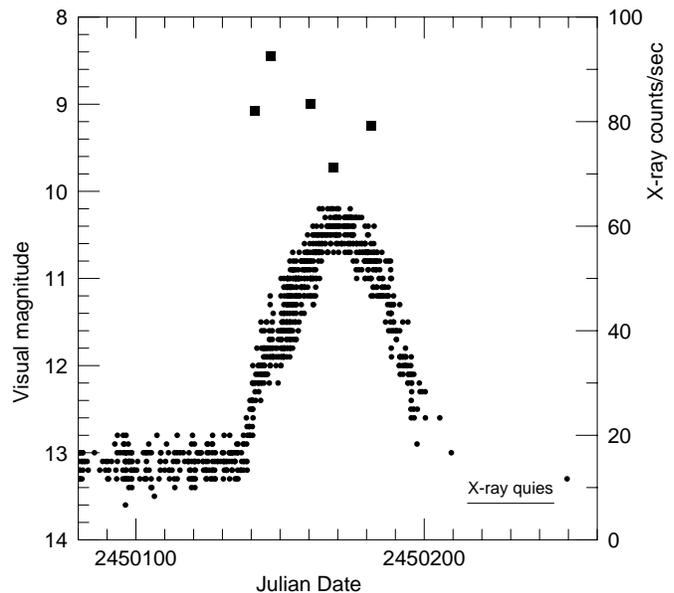} 
\end{figure}

\begin{figure*}\vspace*{10.2cm}     % Fig 2 
\caption{Two samples of the 2--15-\kev\ X-ray lightcurve of GK~Per 
during outburst, in 16-s bins. Errors are typically 3 counts s\minone.
Time zero corresponds to JD(TDB) 2450140.9392 (top panel) and 
2450181.6800 (lower panel).}
\includegraphics{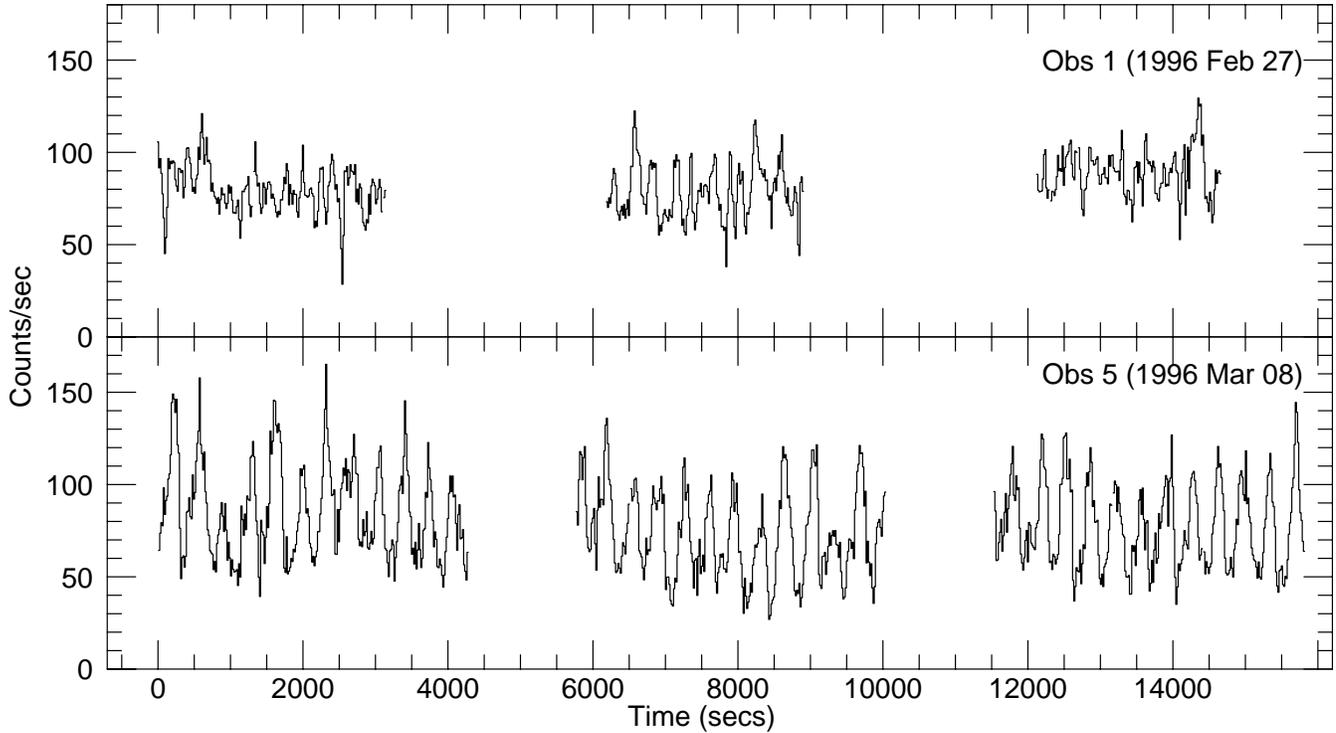} 
\end{figure*}

The mean 2--15-\kev\ flux in the 5 observations was 50\,$\pm$\,8
\pten{-11} ergs\,s\minone\,cm\mintwo; for comparison, Ishida \etal\ 
(1992) report that in {\sl Ginga\/} observations of the 1989 outburst
the 2--20-\kev\ flux was 48\,$\pm$\,17 \pten{-11}\ while the quiescent
flux is much lower at 4.0\,$\pm$\,0.4 \pten{-11}\
erg\,s\minone\,cm\mintwo. 

We show in Fig.~2 sections of lightcurve from the 1\up{st}\ and
5\up{th}\ observations.  The 5\up{th}\ observation shows the
single-peaked, quasi-sinusoidal modulation that is typical of
intermediate polars.  In contrast, the first observation shows a
shallower, flat-topped modulation that is similar to the quiescent
pulse recorded by Ishida \etal\ (1992), despite the count rate being
already well above the quiescent level.

For further comparison we plot the folded pulse profiles of all 5
observations in Fig.~3, and their 6--10/2--5-\kev\ hardness ratios in
Fig.~4.  Note that observations 3 and 4, nearest the peak of the
outburst, have the highest hardness ratios. This can be seen
from Fig.~4 where these observations are displaced upwards from
equal spacing. Spectral analysis confirms that all the outburst
spectra are heavily absorbed, requiring at least two partial-covering
absorbers of densities $\approx$\,3\pten{23}\ cm\mintwo\ and
$\approx$\,2\pten{24}\ cm\mintwo.  This was also reported by Ishida 
\etal\ (1992) during the 1989 outburst, 
and contrasts with typical quiescent 
absorption of only $\approx$\,1\pten{22}\ cm\mintwo. 

Intermediate polar spectra typically show increased hardness in
the minima of the spin pulses, owing to increased absorption, and this
is seen in observations 2 to 5. In contrast, the first observation
shows a more complex and shallower spectral change.

\begin{figure}\vspace*{13.3cm}     % Fig 3 
\caption{The spin-pulse profiles from the 1st (bottom) to 5th (top)
X-ray observations. We have added 50 counts s\up{--1}\ successively to
the upper 4 profiles. Photon-noise error bars are typically 0.5 counts
s\up{--1}.  The data are folded on a period of 351.335 s with phase zero 
corresponding to JD(TDB) 2450168.14105.}
\includegraphics{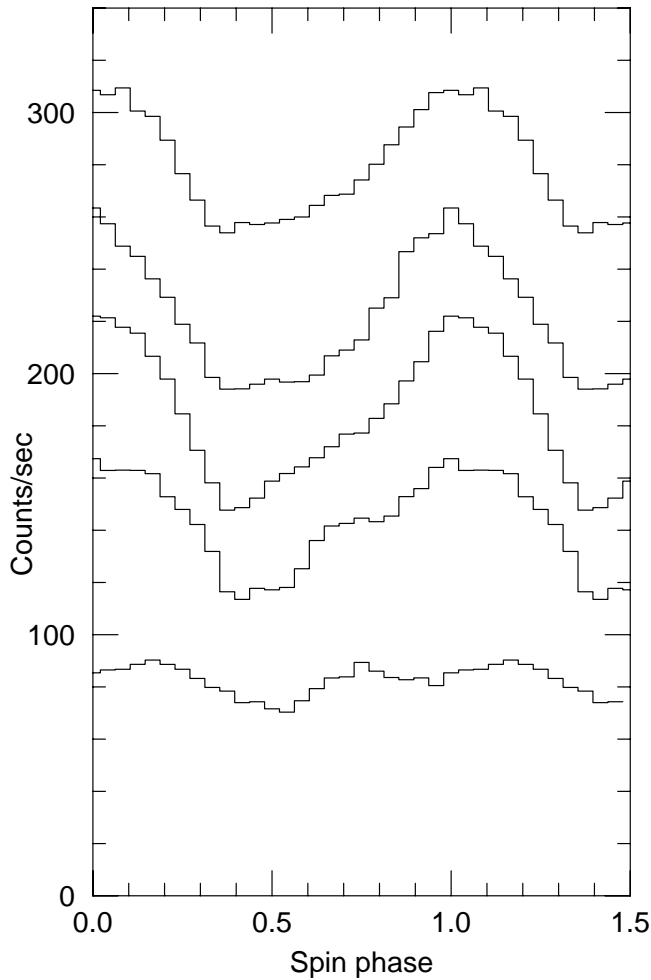} 
\end{figure}

We have used the last 4 observations, which span 35 d, to measure 
the pulsation period (the first was omitted owing to the different
pulse shape).  From these we find a period of 351.335\,$\pm$\,0.002 s, 
where the error makes no allowance for possible changes in pulse profile.
This result is in line with Mauche's (2003) re-assessment of GK~Per's 
period change over that originally proposed by Patterson (1991).

\begin{figure}\vspace*{13.3cm}     % Fig 4 
\caption{The 6--10/2--5-\kev\ hardness ratios, folded on the 
spin cycle, from the 5 \xte\ observations. We have added 0.5 successively 
to the upper 4 profiles. Photon-noise error bars are always less than 0.1 
(typically 0.05). The phase zero is the same as in Fig.~3.}
\includegraphics{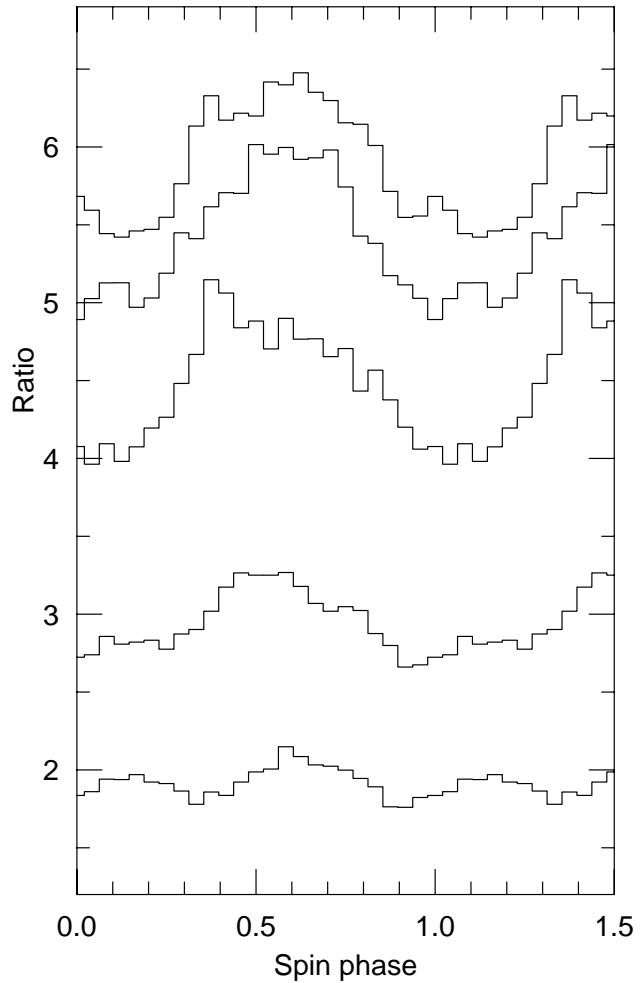} 
\end{figure}

\section{The 5000-s QPOs}
In the {\sl EXOSAT\/} observations of the 1983 outburst Watson \etal\
(1985) reported modulations of the X-ray flux which had a strong
amplitude of up to a factor 2--3, and a quasi-periodic timescale of
\sqig 5000 s. Hellier \&\ Livio (1994) proposed that they were caused
by bulges in the inner disc, orbiting with the local Keplerian
timescale of 5000 s, as might arise from an overflowing stream
reimpacting the disc.  These bulges would periodically obscure the
line of sight to the X-ray emission, causing absorption dips. The dips
are deeper at lower energies, in keeping with this idea.

Morales-Rueda, Still \&\ Roche (1999) analysed optical spectroscopy of
the 1996 outburst and found the same 5000-s QPOs in the emission-line
profiles.  They proposed an alternative model in which blobs at the
inner disc edge orbit with a period of either 320 or 380 s.  These
would cause enhanced flow to the accretion curtains and hence enhanced
absorption whenever a blob lined up with the magnetic dipole, thus 
producing a modulation at the 5000-s beat period between the blob
orbits and the 351-s spin period.

While this idea adequately explains the optical data analysed by
Morales-Rueda \etal\ (1999), it is less able to explain the X-ray
behaviour. For example, the extra absorption would not occur when the
blob-fed curtain was on the far side of the white dwarf, as it would be
for half of each spin cycle. Indeed, one might expect the extra
accretion flow of a blob-fed curtain to result in enhanced flux for
these spin phases. The observations, though, show that the extra
absorption of the QPO occurs throughout the spin cycle, reducing the
flux at all spin phases (see Watson \etal\ 1985 and Hellier \&\ Livio
1994). For this reason we prefer models in which the structures
causing the dips are circling at the 5000-s quasi-periodicity. 

More recently, Warner \&\ Woudt (2002) have proposed a new
understanding of the QPOs and dwarf-nova oscillations (DNOs) seen in
cataclysmic variables. They associate DNOs with a magnetospheric
rotation period and suggest that QPOs are caused by slow-moving
prograde waves at the inner edge of the disc.  Warner, Woudt 
\&\ Pretorius (2003) show that a relation $P_{\rm QPO}/P_{\rm spin} 
\approx 15$ fits many observations in cataclysmic variables and X-ray 
binaries; GK~Per obeys this relation with
$P_{\rm QPO}/P_{\rm spin} \approx 5000/351 \approx 14$.

Given this, we retain Hellier \&\ Livio's (1994) proposal
that GK~Per's QPOs are dipping behaviour caused by bulges moving at
\sqig 5000 s, but now regard Warner \&\ Woudt's (2002) explanation of
the bulges as slow, prograde travelling waves as the most promising.

Morales-Rueda \etal\ (1999) observed that the emission-line QPOs were
predominantly blueshifted, and suggested that this argues against a
model in which structure moves at 5000 s.  However, it is plausible that
the waves have a leading-edge/trailing-edge asymmetry in emissivity,
which would explain the Morales-Rueda \etal\ result.

The relevance to this paper is that optical observations by 
Morales-Rueda \etal\ (1999) and Nogami, Kato \&\ Baba (2002) 
show that the QPOs were present during the 1996 outburst
discussed here.  (We have attempted to detect them in the \xte\ data,
but the search was inconclusive since the data are broken up by \xte's
orbit on the very similar timescale of 6000 s.) Their existence shows
that GK~Per has a sufficiently high inclination that, at least during
outburst, bulges of material at the inner disc edge are capable of
obscuring the line of sight to the white dwarf and reducing the flux
by factors of 2--3. This is important for the discussion on
the origin of the spin pulse.

\begin{figure*}\vspace*{5cm}     % Fig 5 
\caption{An illustration of the change in accretion geometry 
during outburst, showing the magnetosphere shrinking markedly 
as the disc pushes inwards. The dotted lines show the 50--73$^{\circ}$
allowed inclination range. The arrows show how the line-of-sight through
the accretion regions can become more tangential during outburst.}
\includegraphics{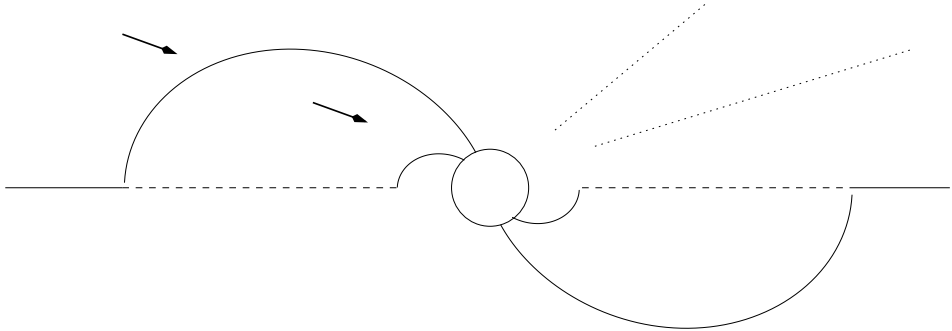} 
\end{figure*}

\section{Comparison with XY~Ari: hiding the lower pole?}
Like GK~Per, XY~Ari shows a low-amplitude, non-sinusoidal pulse in
quiescence.  This most likely results from minor asymmetries between
two accreting poles, so that the disappearance of one as the white
dwarf spins is not entirely balanced by the appearance of the
other. %It also suggests that absorption, occurring when the 
%accretion curtains sweep across the line of sight, has a minor effect
%on the spin pulse. 

In outburst, XY~Ari's pulse changed to a very deep, quasi-sinusoidal
modulation. The minima were not accompanied by increased hardness
suggesting that absorption played no role.  Noting that XY~Ari was
deeply eclipsing and thus at a very high inclination, Hellier \etal\
(1997), suggested that the inner disc had pushed inwards during
outburst, cutting off the line of sight to the lower accreting
pole. Thus we saw only the upper pole, and this passed from the
visible face to the far side and back as the white dwarf rotated,
producing a near-total modulation.

GK~Per's pulse also changes from low-amplitude and non-sinusoidal in
quiescence to deep and sinusoidal in outburst (Fig.~3). It is thus
natural to ask whether the same effect is occurring. An immediate
counter is that, unlike in XY~Ari, GK~Per's outburst pulse is energy
dependent and caused primarily by an increase in absorption at spin
minimum.  Thus the model cannot apply straightforwardly.

Nevertheless, let us consider whether the system parameters are
compatible with hiding the lower pole in outburst.  We assume
a 1-M$_{\sun}$ white dwarf (see Morales-Rueda \etal\ 2002), which,
with a 351-s spin pulse, produces a magnetosphere of \appro 10\,\rwd, 
assuming that it corotates with a Keplerian inner disc.
There are arguments for a much larger magnetosphere, based on 
spectra and modelling the outbursts (e.g.\ Kim \etal\ 1992), but
the presence of pulsed X-rays in quiescence implies that 
accretion is not centrifugally prohibited, so we place the 
inner disc edge at the corotation radius.

A linear scaling with optical and X-ray brightness would suggest that
the accretion rate rises by a factor \appro 10 during outburst.
However, this is not in accord with disc-instability models of
the outburst, and Kim \etal\ (1992) and Yi \etal\ (1992) argue that
the accretion rate rises by a factor \sqig 100--1000, with the X-rays
being supressed by opacity.  The standard scaling of disc radius with
accretion rate ($r\!\propto\!\dot{m}^{-2/7}$) then implies that the
inner disc radius decreases by a factor of between 4 and 7 (we show an
illustrative reduction by a factor 4 in Fig.~5). The system 
inclination is within the range 50--73$^{\circ}$ (Morales-Rueda
\etal\ 2002). 

The above ranges are not tight enough to tell
us whether the bottom of the white dwarf is hidden by the disc
in outburst (particularly given the uncertainty in the disc
thickness), but allow both possibilities.  
We thus turn to considering the observations.

The high amplitude of the 5000-s QPOs (reductions in X-ray flux by
factors up to 2--3) imply that bulges near the inner-disc edge  
obscure the upper accretion pole, and are thus of a height
above the disc plane comparable to the white-dwarf radius. 
This is reinforced by the observation of very high absorption
(10$^{24}$ cm\mintwo) at all times during outburst. Given the
inclination range of 50--73$^{\circ}$, and the likely
inner disc radius of $<$\,3\,\rwd, such material would likely
hide the lower pole continually. 

Note, however, that we don't see episodes of near-zero flux each
spin cycle, which we
do in XY~Ari when the upper pole swings round to the hidden face
of the white dwarf. This implies either that the lower pole is visible
at phases when the upper pole disappears, or that accretion
regions at the upper pole are visible at all spin phases.
The latter is, at first sight, implausible for the above inclination
range.  The small magnetosphere during outburst implies accretion
regions far from the poles ($>$\,35$^{\circ}$ magnetic co-latitude
for $r_{\rm mag}$\,$<$\,3$R_{\rm wd}$), which, added to any dipole
offset from the spin axis, will likely exceed the $90^{\circ}$$-$\,$i$ 
polecap region that is always visible. 

So can we reconcile these conflicting indicators?  We suggest that,
in outburst, the accretion flow overwhelms the magnetosphere sufficiently
that accretion flows to the poles from all azimuths (whereas, in
quiescence, each pole would be fed from a restricted azimuthal range), 
and thus falls at all magnetic longitudes. Hence accretion
at the upper pole would always be visible.   There is evidence 
from eclipse timings that exactly this occurs at the peak of
the outburst in XY~Ari (Hellier \etal\ 1997).  If the above is correct,
it suggests that the angle between the spin and magnetic axes is
relatively small. 

\section{The change in the spin pulse}
The fact that the pulse profile in GK~Per is 
largely an absorption dip, and not the near-total, energy-independent
modulation seen in XY~Ari in outburst, implies that simply hiding
the lower pole does not explain the change in pulse profile
between quiescence and outburst. We are thus left to explain
the fact that (1) in outburst GK~Per shows a `typical' pulsation with a
quasi-sinusoidal profile resulting from a broad absorption dip,
and (2) the absorption dip is much reduced in quiescence leaving a 
more-complex, lower-amplitude modulation. 

We suggest that the change results straightforwardly from a
combination of three factors. First, the increase in accretion rate by
a factor \sqig 100--1000 will increase the column density of the
accretion curtains. Second, the marked shrinking of the accretion
curtains in outburst (see Fig.~5) will force the increased flow
through a much smaller circumference and thus a much smaller 
area, further increasing the column
density. Third, as illustrated in Fig.~5, the change in geometry
can easily result in the line-of-sight to the accretion regions taking
a more grazing path through the accretion curtains, thus increasing
the line-of-sight column yet further.  Such effects can explain why we
observe a phase-varying column as high as 2\pten{24}\ cm\mintwo\ in
outburst, whereas the quiescent spin pulse is modelled by phase-varying
absorption of only 6\pten{21}\ cm\mintwo\ (Ishida \etal\
1992).

Thus, in outburst, the effect of intense absorption dominates the pulse
profile, whereas in quiescence the much weaker absorption, from
tenuous accretion curtains much further from the white dwarf, is
less significant than other factors, such as an asymmetry between the
two poles. 

\section{Conclusions}
\hspace*{5mm}
(1) X-ray QPOs are caused by bulges at the inner disc edge, travelling
at a 5000-s period.  They likely correspond to the travelling waves
discussed by Warner \&\ Woudt (2002).  The bulges are at a height
above the disc plane comparable to the white-dwarf radius.

(2) We suggest that the X-ray QPOs are not seen in quiescence
owing to the inner disc being much further out, so that the
bulges do not obscure the white dwarf.

(3) During outburst, accretion occurs from all azimuths, forming
a complete accretion ring at the poles.  The lower pole is
likely hidden.

(4) The pulse profile changes in outburst to become dominated by 
absorption; this results from the greater accretion flow,
the smaller accretion-curtain area, and the change in how the
line-of-sight passes through the curtains. 

(5) Measurement of the 351-s spin period confirms the period
change reported by Mauche (2003). 

\section*{Acknowledgments}
We thank Janet Mattei and the many observers who contribute
to the AAVSO for compiling and supplying visual observations 
of GK~Per's outburst.

\end{document}